\documentstyle[12pt]{article}
\begin{document}
\def\la{\langle}
\def\ra{\rangle}
\def\wt{\widehat}
\def\o{\overline}
\newcommand{\beq}{\begin{equation}}
\newcommand{\eeq}{\end{equation}}
\newcommand{\beqa}{\begin{eqnarray}}
\newcommand{\eeqa}{\end{eqnarray}}
\begin{center}
{\Large\bf Average local values and local variances in
quantum mechanics 
\vspace*{1.1cm}\\}
{\large J.G. Muga, J.P. Palao and R. Sala\vspace*{.3cm}\\}
{\it Departamento de F\'{\i}sica Fundamental
y Experimental, Facultad de F\'{\i}sica, Universidad de
La Laguna, Tenerife, Spain}
\end{center}

\pagestyle{plain}
\baselineskip 16pt
\vspace*{1.2cm}
\begin{center}
{\bf Abstract}
\end{center}
Several definitions for the average local value and 
local variance of a quantum observable are examined
and compared with their classical counterparts. 
An explicit way to construct an infinite number of these 
quantities is provided. It is found that different classical
conditions may be satisfied by different definitions, but 
none of the quantum definitions examined
is entirely consistent with all classical
requirements.  
$^{}$\vspace*{1.4cm}\\
{\small  Ref: Physics Letters A 238(1998)90, Electronic version with permission of Elsevier\\
e-mail address: JMUGA@ULL.ES\\
Fax number: 34-22-603684
\vspace*{.2cm}}  
\newpage
\baselineskip 22pt
Many simple concepts of standard classical statistical mechanics
are not easy to translate into quantum mechanics. The average local
value and the local spread of a dynamical variable belong to this
group. For an ensemble of particles in one dimension characterized
by a joint distribution $F(q,p)$ of position $q$ and momentum $p$ the
average local value of the dynamical variable $a(q,p)$ is given by 
\beq\label{aq}
\o{a}(q)=\frac{1}{P(q)}
\int\!\!\int  a(q',p)\delta(q'-q) F(q',p)\,dq' dp,
\eeq
where $P(q)\equiv\int F(q,p)\,dp$, 
and the local spread, or local variance, by 
\beq\label{saq}
\sigma^2_{a|q}=\overline{a^2}(q)-[\bar{a}(q)]^2\,,
\eeq
where $\o{a^2}$ is the local average of the square of $a$, 
or second order ``local moment''. 
Of course an arbitrary power of $a$, $b(q,p)\equiv a^n(q,p)$,
is also a
function of $(q,p)$, so the expression (\ref{aq}),
mutatis mutandis, gives also 
the local moments of arbitrary order,   
\beqa\label{mn}
\bar{b}(q)&=&\frac{1}{P(q)}
\int\!\!\int  b(q',p)\delta(q'-q) F(q',p)\,dq' dp
\\
&=&\o{a^n}(q)=\frac{1}{P(q)}\int\!\!\int  a^n(q',p)
\delta(q'-q) F(q',p)\,dq' dp\,.
\eeqa
This observation may appear trivial at this point but it will become 
important when discussing the quantum case. 
Also, the following marginal, joint, and conditional probabilities
can be defined,
\beqa
P(a)&=&\int\!\!\int \delta[a-a(q,p)] F(q,p)\,dq\, dp\\
P(a,q)&=&\int\!\!\int \delta[a-a(q',p)]\delta(q'-q)
F(q',p)\,dq'\,dp\\
P(a|q)&=&P(a,q)/P(q)
\eeqa
where the last equation is simply Bayes' rule.  
In terms of these later quantities $\bar{a}$ and  
$\sigma^2_{a|q}$ take the form
\beq\label{9}
\bar{a}(q)=\int a\,P(a|q)\,da=\frac{1}{P(q)}\int a\, P(a,q)\, da
\eeq
and  
\beq
\sigma^2_{a|q}=\frac{1}{P(q)}\int (a-\bar{a})^2 P(a,q)\,da\,.
\eeq
The total (or global) average is simply the 
``$q$-average'' [i.e. an average over $q$ of a $q$-dependent
function weighted by $P(q)$] of the local averages,
\beq
\la a\ra =\int\!\!\int  a(q,p) F(q,p)\,dq\, dp
=\int \o{a}(q) P(q)\,dq\,,   
\eeq
whereas the total variance takes the form,
\beqa
\sigma^2_a&\equiv&\int\!\!\int \left(a-\la a\ra\right)^2
P(a,q)\,da\,dq\\
&=&  
\int \sigma^2_{a|q} P(q)\,dq  +\int (\o{a}(q)-
\la a\ra)^2 P(q)\,dq\,.
\label{deco}
\eeqa
$\sigma_a^2$ is the $q$-average of the local variances plus 
the ``$q$-variance'' of the local averages. This is an appealing
decomposition, and it seems reasonable to seek for definitions of 
quantum local moments that preserve its structure.      
     
In a recent publication in this journal, L. Cohen [\ref{Cohen96}]
has argued
that it is natural to interpret the following quantities
as the local value of the observable associated with the hermitian
operator 
$\wt{A}$ and its local spread,
\beqa
\label{aqr}
\bar{A}^S(q)&\equiv&\left(\frac{\la q|\wt{A}|\psi\ra}
{\la q|\psi\ra}\right)_{\cal R}
\\
\label{sqi}
[\sigma^2_{A|q}]^C&\equiv&\left(\frac{\la q|\wt{A}|\psi\ra}
{\la q|\psi\ra}\right)^2_{\cal I}\,.
\eeqa
The superscripts $S$ and $C$ will distinguish these quantities
from others to be defined later (the reason for using different
letters will become clear soon), and the subscripts ${\cal R}$ and
${\cal I}$ indicate ``real'' and ``imaginary'' parts. (This
interpretation of (\ref{aqr}) has also been proposed in  
[\ref{Holland}-\ref{Sut}].) These quantities obey a 
relation with the form of Eq. (\ref{deco}),
\beq\label{qu}
\sigma^2_{A}=\int \sigma^2_{A|q} |\psi(q)|^2 dq
+\int (\bar{A}^S(q)-\la\wt{A}\ra)^2 |\psi(q)|^2 dq\,, 
\eeq
and the analogy with the classical equation has been invoked to understand
the significance of the two terms in (\ref{qu}) [\ref{Cohen96}].
In this letter we shall explore how far the classical-quantum analogy goes.    
We shall see in particular that the definitions (\ref{aqr}) and (\ref{sqi}) 
do not always lead to results consistent with the form of
equations (\ref{saq}) and (\ref{mn}).
To this end it is useful to introduce 
an  operator that symmetrizes $\wt{A}$ and 
$\delta(\wt{q}-q)=|q\ra\la q|$,
\beq
\wt{A_q}=\frac{1}{2}
\big[\wt{A}\delta(\wt{q}-q)+\delta(\wt{q}-q)\wt{A}\big]\,,
\eeq
and express (\ref{aqr}) 
in terms of it as, 
\beq
\bar{A}^S(q)=\frac{\la\wt{A_q}\ra}{\varrho(q)}\,.
\eeq
where $\varrho(q)=|\psi(q)|^2$ is the probability density
and $\la\wt{A_q}\ra\equiv\la\psi|\wt{A_q}|\psi\ra$. 
Note that the expectation value of $\wt{A_q}$ is a ``local
density of $A$'', i.e., a quantity that integrated
over $q$ provides the global average,  
\beq
\int \la\psi|\wt{A_q}|\psi\ra \,dq=\la \wt{A}\ra.
\eeq
(We shall see below that other definitions also satisfy this
condition.) 
For consistency, the definition for the local average 
should apply to any operator, and in particular to $\wt{A}^2$. 
The local average of $\wt{A}^2$ is accordingly given by  
\beq
\o{A^2}^{\,S}(q) = \frac{\la(\wt{A}^2)_q\ra}{\varrho(q)}
=\frac{1}{2\varrho(q)}\la\psi|
[\wt{A}^2\delta(\wt{q}-q)+\delta(\wt{q}-q)
\wt{A}^2]|\psi\ra\,.
\eeq
Following (\ref{saq}), a quantum 
local variance for $\wt{A}$ is ``naturally'' defined as 
\beqa\label{new}
[\sigma_{A|q}^2]^S&=&\o{A^2}^S(q)-[\bar{A}^S(q)]^2
\\
&=&\frac{1}{2\varrho(q)}\la\psi|
[\wt{A}^2\delta(\wt{q}-q)+\delta(\wt{q}-q)
\wt{A}^2]|\psi\ra-[\bar{A}^S(q)]^2
\eeqa
where the superscript $S$ reminds that only symmetrized operators 
are used.   
This definition for the local 
variance also satisfies the decomposition (\ref{deco}), but it
is different from (\ref{sqi}).
To see this in more detail let us write (\ref{sqi}) as 
\beqa
[\sigma^2_{A|q}]^C&=&\left(\frac{\la q|\wt{A}|\psi\ra}
{\la q|\psi\ra}\right)^2_{\cal I}=
\left|\frac{\la q|\wt{A}|\psi\ra}
{\la q|\psi\ra}\right|^2-\left(\frac{\la q|\wt{A}|\psi\ra}
{\la q|\psi\ra}\right)^2_{\cal R}
\\
\nonumber
&=&\frac{1}{\varrho(q)}\la\psi|\wt{A}\delta(\wt{q}-q)\wt{A}|\psi\ra
-[\bar{A}^S(q)]^2
\eeqa
Clearly the two procedures imply two different interpretations of the 
local density for the square of $\wt{A}$. In general, 
\beq
\la\psi|\wt{A}\delta(\wt{q}-q)\wt{A}|\psi\ra\ne
\frac{1}{2}\la\psi|[\wt{A}^2\delta(\wt{q}-q)+\delta(\wt{q}-q)
\wt{A}^2]|\psi\ra\,.
\eeq
However their integrals over $q$ are both equal to $\la \wt{A}^2\ra$.  
Eq. (\ref{new}) provides a local variance consistent with the
definition 
given for the local average and it is in agreement with the
classical expressions (\ref{deco}) and (\ref{saq}), {\it but} it is
not semidefinite positive. Its literal interpretation as a variance 
is therefore impossible. This resembles the status of the Wigner
function and other phase space quasi-distribution functions. While
not
interpretable as probability distributions they can be used to
correctly
evaluate expectation values and to investigate the classical limit
in a classical-like phase space language.

In fact the decompositions of the 
total variance compatible with (\ref{deco}) and (\ref{saq}) are infinite. 
They can be found by means of the phase space formalisms
described  within the general framework 
provided also by Cohen [\ref{Cohen},\ref{SPM}].
Each of these formalisms is  associated with 
a particular function $f(\theta, \tau)$ of auxiliary
variables $\theta$ and $\tau$.   
The density operator and the operator $\wt{A}$ are related,
respectively,
to a quasi distribution function $F(q,p;[f])$ and a phase space
representation $\tilde{A}(q,p;[f])$ [both depend functionally on $f$; 
Note that $\tilde{A}(q,p;[f])$ is not necessarily equal to the
classical function $a(q,p)$] in such a way that the expectation
value of $\wt{A}$ is given by the phase space integral,
$\la\wt{A}\ra=\int\!\!\int \tilde{A}(q,p;[f])F(q,p;[f])\, dq dp$.
These formalisms are also closely related to the 
``quantization
rules'', that define mappings from phase
space functions to operators. The rules are
generally
used to associate a quantum operator with the classical function
$a(q,p)$.  

Local values can be defined using any of these 
phase space formalisms, see the classical expression (\ref{9}), as        
\beq\label{laf}
\bar{A}(q;[f])=\frac{1}{\varrho(q)}\int \tilde{A}(q,p;[f])
F(q,p;[f])\,dp\,.
\eeq
Their $q$-average is the global average,
$\la\wt{A}\ra=\int\bar{A}\,\varrho\, dq$,
and the variance can be decomposed in agreement
with (\ref{deco}) by writing 
\beqa
\nonumber
\sigma_A^2
&=&\int \bigg[\int \frac{\widetilde{A^2}(q,p;[f])
F(q,p;[f])}{\varrho(q)}\,dp\bigg]
\varrho(q) dq
-\int \bar{A}^2(q;[f])\ \varrho(q)\,dq
\\
&+&\int \bar{A}^2(q;[f]) \varrho(q)\,dq-\la \wt{A}\ra^2\,,
\label{24}
\eeqa
where $\widetilde{A^2}(q,p;[f])$ is the phase space representation
of $\wt{A}^2$.   

The first two terms in (\ref{24}) may be 
regarded as the $q$-average of the ``local variance''
\beq\label{lvf}
\sigma_{A|q}^2[f]\equiv \o{A^2}(q;[f])
-\bar{A}^2(q;[f])\,,
\eeq
while the last two terms take the form of a $q$-variance
of local averages.
An important point is that $\widetilde{A^2}(q,p;[f])$ is in general different
from the square $\tilde{A}^2(q,p;[f])$ so it is not necessarily a
positive function. Unless $f$ is suitably chosen $F$ is not positive either, 
so that, contrary to the classical local variance, and in spite of the 
``square'' used in the notation, $\sigma_{A|q}^2[f]$
is not necessarily positive. Of course for a given kernel $f$,
$\sigma_{A|q}^2[f]$ can be positive for particular observables and,
similarly, for a given observable a family 
of kernels will give a positive local variances. In the case
of momentum, $\wt{A}=\wt{p}$, Cohen has noted that (\ref{lvf}) is positive
for a family of kernel functions $f$ that lead precisely to the choices
(\ref{aqr}) and (\ref{sqi}) [\ref{Cohen90}].
In general, however, each observable requires a different analysis.    
     
We shall next elaborate on the momentum observable and its powers using   
two of these formalisms because of their close relation to the previously
discussed choices and to classical expressions.     
The Rivier-Margeneau-Hill (RMH) formalism is obtained by 
taking $f=\cos(\theta\tau\hbar/2)$
[\ref{Rivier},\ref{MarHill},\ref{Cohen},\ref{SPM}].
We shall denote the corresponding representation of the state,
namely the 
Margeneau-Hill function, as $F^{MH}$. The subscript $MH$ will be also
used for the local averages and variances, (\ref{laf}) and
(\ref{lvf}),
defined within this framework. The corresponding
mapping from phase space representatives to operators is the         
``Rivier rule'' [\ref{Rivier}].
When applied to a phase space function 
with the factorized form $g(q)h(p)$ it gives the symmetrized
operator $[g(\wt{q})h(\wt{p})+h(\wt{p})g(\wt{q})]/2$. 
Note that for an arbitrary product of phase space functions  $AB$, this
symmetrization is not always equal to the simple symmetrization rule
$(\wt{A}\wt{B}+\wt{B}\wt{A})/2$, where $\wt{A}$ and $\wt{B}$ are quantum
operators assigned to $A$ and $B$ by some prescription (that could be
in fact Rivier's rule), see e.g. [\ref{GG}].
However, if $A(q)$ and $B(p)$ are, respectively, 
functions of $q$ and $p$ only, and the associated operators are  
$A(\wt{q})$ and $B(\wt{p})$, then the two symmetrization
procedures agree. This implies in particular that
$\o{p^n}^S(q)=\o{p^n}^{MH}(q)$ and $[\sigma_{p|q}^2]^S=
[\sigma_{p|q}^2]^{MH}$.  

It has been argued that 
the interpretation of (\ref{aqr}) as the local average leads
to the Margeneau-Hill function [\ref{Sut}].
This connection can indeed be made, but, remarkably,  
the local values obtained by means of
the RMH formalism in general differ from (\ref{aqr}). To understand
how this may happen let us briefly 
review the derivation of the Margeneau-Hill function in [\ref{Sut}].        
The idea is to define a conditional probability density 
$P^S(p|q)$ making use of the characteristic function concept, i.e., 
by inverting 
\beq
G(\tau,q)\equiv \o{e^{i\tau p}}^S(q)=\int e^{i\tau p} P^S(p|q)\,dp      
\eeq
\beq
P^S(p|q)\equiv\frac{1}{2\pi}\int e^{-i\tau p}\, \o{e^{i\tau p}}^S(q)\,
d\tau
\eeq
Expanding the exponential in $\o{e^{i\tau p}}^S$, there results a
series where 
the coefficients are the local averages of the powers of $p$,
$\o{p^n}^S$.
Using the coordinate representation for the momentum operator the
Taylor series for $\psi(q\pm\hbar\tau)$ can be recognized,
\beq       
\o{e^{i\tau p}}^S(q)=\sum_n \frac{(i\tau)^n}{n!} \o{p^n}^S(q)
=\frac{\psi(q+\hbar\tau)}{2\psi(q)}+
\frac{\psi^*(q-\hbar\tau)}{2\psi^*(q)}\,.
\eeq
If in addition one {\it defines} a quasi-joint probability 
distribution by
the product $\varrho(q)P^S(p|q)$, i.e., by 
formally adopting the structure of Bayes' theorem,  
this quasi-probability 
turns out to be the Margeneau-Hill function,
\beq\label{MH}
F^{MH}(q,p)=[\la p|\psi\ra\la\psi|q\ra\la q|p\ra]_{\cal R}\,.
\eeq
But, as discussed before, in general
$\bar{A}^S\ne \bar{A}^{MH}$. The symmetrized
operator $(\wt{A}^2)_q$ is equal to the one provided by 
Rivier's rule if $A$ is only a function of $p$, but will differ in 
general. The use of the structure of Bayes' theorem in the 
derivation of (\ref{MH}) is the key point that explains
this seeming contradiction. This is a theorem valid for
actual  probabilities, but $P^S(p|q)$ and $F^{MH}(q,p)$ are not.
In fact they can be negative.  
 
Finally, the comparison with the classical equations satisfied
by local 
average and the variance is completed here by studying the time
dependence, in particular the equations for time
derivatives of the local averages of the powers of $p$, i.e., the
equations of hydrodynamics. In this context the Weyl-Wigner
(WW) phase space formalism  is privileged.
This formalism is associated with
the simplest choice, $f=1$ [\ref{WW},\ref{Cohen},\ref{SPM}].
The phase space representative of the state
is the Wigner function, $F^W(q,p)$, and the corresponding rule is
the Weyl quantization rule. A superscript $W$ will denote the local
quantities (\ref{laf}) and (\ref{lvf})
calculated with this formalism. 

Cohen's prescription, Eq. (\ref{aqr}),
and the phase space formalisms RMH and WW lead to the 
same equation for the local density,
namely the continuity equation,
\beq
\frac{\partial\varrho(q)}{\partial t}
=-\frac{\partial}{\partial q}[\varrho(q)\bar{p}(q)]\,,
\eeq
where $\bar{p}=\bar{p}^S=\bar{p}^{MH}=\bar{p}^W$.
For the first local moment, the Wigner function leads to an 
``equation of motion'' with exactly the same {\it form}
as the classical one [\ref{PS}], 
\beq
\frac{\partial\bar{p}^{W}}{\partial t}=
-\frac{\bar{p}^W}{m}\frac{\partial \o{p}^W}
{\partial q}-\frac{\partial{V(q)}}{\partial q}
-\frac{1}{m\varrho(q)}
\frac{\partial\big(\varrho(q)[\sigma^2_{p|q}]^W\big)}{\partial q}, 
\eeq
$m$ being the mass, and $V(q)$ the potential function.
(Even though the form is equal, the numerical values differ
in general between the classical and the quantum cases, 
and the local variance $[\sigma_{p|q}^2]^W$ is not
semidefinite positive [\ref{PS}].)  
However, in the other approaches the local variances are 
different,
\beqa
[\sigma_{p|q}^2]^W&=&[\sigma_{p|q}^2]^{MH}+
\frac{1}{4\varrho(q)}\la 2\wt{p}\,\delta(\wt{q}-q)\wt{p}-\wt{p}\,
\delta(\wt{q}-q)-\delta(\wt{q}-q)
\wt{p}\ra
\\
&=&[\sigma_{p|q}^2]^{C}-
\frac{1}{4\varrho(q)}\la 2\wt{p}\,\delta(\wt{q}-q)\wt{p}-\wt{p}\,
\delta(\wt{q}-q)-\delta(\wt{q}-q)
\wt{p}\ra\,,
\eeqa 
and there appear   
extra terms which are not present in the classical
equation (for the RMH formalism they were studied
by Sonego [\ref{Sonego}]). 
At least for several model potentials studied in ref. [\ref{RMB}] 
the WW formalism is not only closer to classical
mechanics {\it formally} in this context; it is also numerically
closer. The local kinetic energy densities 
$\varrho(q)\o{p^2}^W(q)/(2m)$, $\varrho(q)\o{p^2}^{MH}(q)/(2m)$, and
$\la \wt{p}\delta(\wt{q}-q)\wt{p}\ra/(2m)$
were evaluated, and the one provided by the WW formalism 
was clearly the closest (numerically) to
the classical values [\ref{RMB}].
Determining the extent of this agreement is an 
interesting open question for a separate study. 

In summary, we have examined different definitions of ``quantum local
averages and variances'', and their similarities and
differences with various classical
expressions. None of them satisfies all the relations valid
classically. It is useful to maximize the agreement
with classical mechanics for examining the classical limit,
and to gain physical insight in certain applications,
but it should be noted
that different classical criteria are satisfied by
different quantum definitions. 
For specific applications one of the definitions may turn out 
to be the most convenient, but in fact each of
them contains a piece of information which is only partially
analogous to the corresponding classical quantity. 
 
{\bf Acknowledgments}

We acknowledge the referee for very interesting comments. 
Support by Gobierno Aut\'onomo de Canarias (Spain) (Grant PI2/95)
and by Ministerio de Educaci\'on y Ciencia (Spain) (PB 93-0578)
is acknowledged. JPP acknowledges an FPI fellowship from Ministerio 
de Educaci\'on y Ciencia.   

\newpage

\centerline{\large{\bf References}}

\begin{enumerate}

\item\label{Cohen96} L. Cohen, Phys. Lett. A 212 (1996) 315

\item\label{Holland} P. R. Holland, The quantum theory of motion (Cambridge
Univ. Press, Cambridge, 1993)

\item\label{WS} K. K. Wan and P. Summer, Phys. Lett. A 128 (1988) 458

\item\label{Sut} R. I. Sutherland, J. Math. Phys.
23 (1982) 2389
 
\item\label{Cohen} L. Cohen, J. Math. Phys. 7 (1966) 781. 

\item\label{SPM} R. Sala, J. P. Palao and J. G. Muga, 
Phys. Lett. A, accepted 
	
\item\label{Cohen90} L. Cohen, Found. Phys. {\bf 20} (1990) 1455;
L. Cohen, Time-Frequency analysis (Prentice-Hall, Englewood Cliffs,
New Jersey, 1995) 

\item\label{Rivier} D. C. Rivier, Phys. Rev.  83 (1957) 862.

\item\label{MarHill} H. Margenau  and R. N. Hill, 
Progr. Theoret. Phys. (Kioto) 26 (1961) 722;
G. C. Summerfield and P. F. Zweifel, J. Math. Phys. 
10 (1969) 233

\item\label{GG}E. T. Garc\'\i a \'Alvarez and A. D. Gonz\'alez,
Am. J. Phys. 59 (1991) 279

\item\label{WW} E. Wigner, Phys. Rev. A 40 (1932) 749;
J. E. Moyal,  Proc. Cambridge Philos.
Soc.  45 (1949) 99;
H. Weyl, The Theory of Groups and Quantum
Mechanics (Dover, New York, 1950)

\item\label{PS} J. G. Muga, R. Sala and R. F. Snider, 
Physica Scripta 47 (1993) 732

\item\label{Sonego} S. Sonego, Phys. Rev. A
42 (1990) 3733

\item\label{RMB} C. C. Real, J. G. Muga and S. Brouard, 
Am. J. Phys. 65 (1997) 157
\end{enumerate}
\end{document}